\documentclass{aa}
\usepackage{color}
\usepackage{graphicx}
\usepackage{lscape}
\usepackage{natbib}
\usepackage[scaled]{helvet}
\usepackage[varg]{txfonts}
\usepackage{url}
\usepackage{xspace}
\bibpunct{(}{)}{;}{a}{}{,}
\newcounter{Rco}
\newcommand{\Ionst}[1]{\setcounter{Rco}{#1}\Roman{Rco}}

\newcommand{\Ionw}[3]{\mbox{#1\,{\scriptsize\Ionst{#2}}~$\lambda\,#3$\,\AA}\xspace}

\newcommand{\Ionww}[3]{\mbox{#1\,{\scriptsize\Ionst{#2}}~$\lambda\lambda\,#3$\,\AA}\xspace}

\newcommand{\logg}{\mbox{$\log g$}\xspace}
\newcommand{\loggw}[1]{\mbox{$\log g\hspace{-0.5mm} =\hspace{-0.5mm}  #1$}}

\newcommand{\sla}{\raisebox{-0.10em}{$\stackrel{<}{{\mbox{\tiny $\sim$}}}$}}

\newcommand{\Teff}{\mbox{$T_\mathrm{eff}$}\xspace}
\newcommand{\Teffw}[1]{\mbox{$\Teff\hspace{-0.5mm} =\hspace{-0.5mm} #1 \,\mathrm{K}$}}

\newcommand{\Msol}{$M_\odot$\xspace}
\newcommand{\gb}{\object{G191$-$B2B}\xspace}
\newcommand{\re}{\object{RE\,0503$-$289}\xspace}
\begin{document}
\title{Stellar laboratories}
\subtitle{III. New \ion{Ba}{v}, \ion{Ba}{vi}, and \ion{Ba}{vii} oscillator strengths and the barium abundance \\
              in the hot white dwarfs \gb and \re
           \thanks
           {Based on observations with the NASA/ESA Hubble Space Telescope, obtained at the Space Telescope Science 
            Institute, which is operated by the Association of Universities for Research in Astronomy, Inc., under 
            NASA contract NAS5-26666.
           }$^,$
           \thanks
           {Based on observations made with the NASA-CNES-CSA Far Ultraviolet Spectroscopic Explorer.
           }$^,$
           \thanks
           {Tables 1, 2, and 3 are only available at the CDS via anonymous ftp to
            cdsarc.u-strasbg.fr (130.79.128.5) or via
            http://cdsarc.u-strasbg.fr/viz-bin/qcat?J/A+A/vol/page
           }
         }
\titlerunning{Stellar laboratories: new \ion{Ba}{v}, \ion{Ba}{vi}, and \ion{Ba}{vii} oscillator strengths}

\author{T\@. Rauch\inst{1}
        \and
        K\@. Werner\inst{1}
        \and
        P\@. Quinet\inst{2,3}
        \and
        J\@. W\@. Kruk\inst{4}
        }

\institute{Institute for Astronomy and Astrophysics,
           Kepler Center for Astro and Particle Physics,
           Eberhard Karls University,
           Sand 1,
           72076 T\"ubingen,
           Germany \\
           \email{rauch@astro.uni-tuebingen.de}
           \and
           Astrophysique et Spectroscopie, Universit\'e de Mons -- UMONS, 7000 Mons, Belgium
           \and
           IPNAS, Universit\'e de Li\`ege, Sart Tilman, 4000 Li\`ege, Belgium
           \and
           NASA Goddard Space Flight Center, Greenbelt, MD\,20771, USA}

\date{Received 25 March 2014; accepted 19 April 2014}

\abstract {For the spectral analysis of high-resolution and high-signal-to-noise (S/N) spectra of hot stars,
           state-of-the-art non-local thermodynamic equilibrium (NLTE) 
           model atmospheres are mandatory. These are strongly
           dependent on the reliability of the atomic data that is used for their calculation.
          }
          {Reliable \ion{Ba}{v-vii} oscillator strengths 
           are used to identify Ba lines in the spectra of 
           the DA-type white dwarf \gb and 
           the DO-type white dwarf \re and
           to determine their photospheric Ba abundances.
          }
          {We newly calculated \ion{Ba}{v-vii} oscillator strengths
           to consider their radiative and collisional bound-bound transitions
           in detail in our NLTE stellar-atmosphere models
           for the analysis of Ba lines exhibited in
           high-resolution and high-S/N UV observations of \gb  and \re.
          }
          {For the first time, we identified highly ionized Ba in the spectra of
           hot white dwarfs.
           We detected \ion{Ba}{vi} and \ion{Ba}{vii} lines in the 
           Far Ultraviolet Spectroscopic Explorer (FUSE) spectrum
           of \re. The \ion{Ba}{vi} / \ion{Ba}{vii} ionization equilibrium is well
           reproduced with the previously determined effective temperature of 70\,000\,K
           and surface gravity of \loggw{7.5}. The Ba abundance is $3.5 \pm 0.5 \times 10^{-4}$
           (mass fraction, about 23\,000 times the solar value).
           In the FUSE spectrum of \gb, we identified the strongest \ion{Ba}{vii} line (at 993.41\,\AA) only,
           and determined a Ba abundance of $4.0 \pm 0.5 \times 10^{-6}$ (about 265 times solar).
          }
          {Reliable measurements and calculations of atomic data are a pre-requisite for
           stellar-atmosphere modeling. 
           Observed \ion{Ba}{vi-vii} line profiles in two white dwarfs' (\gb and \re) 
           far-ultraviolet spectra were well reproduced with our newly calculated oscillator strengths. 
           This allowed to determine the photospheric Ba abundance of these two stars precisely.
          }


\maketitle

\section{Introduction}
\label{sect:intro}

In recent analyses of 
the hydrogen-rich DA-type white dwarf (WD) 
\gb \citep[effective temperature \Teffw{60\,000}, 
surface gravity $\log\,(g\,/\,\mathrm{cm/s^2}) = 7.6$,][]{rauchetal2012,rauchetal2013,rauchetal2014} and
the hydrogen-deficient DO-type WD 
\re \citep[\Teffw{70\,000}, \loggw{7.5}][]{werneretal2012,rauchetal2014}, numerous lines of the
trans-iron elements Zn, Ga, Ge, As, Se, Kr, Mo, Sn, Te, I, and Xe were identified. This substantially
reduced the number of unidentified lines in the spectra of these two WDs. For precise abundances
determinations, reliable transition probabilities are mandatory -- these are necessary not only for the 
identified lines themselves but for the complete model atom that is considered in the model-atmosphere and
spectral-energy-distribution (SED) calculations. Thus, abundance determinations were so far restricted to
Zn \citep{rauchetal2014},
Ge \citep{rauchetal2012},
Kr, Xe \citep{werneretal2012}, and
Sn \citep{rauchetal2013}.

A close inspection of the still unidentified lines in the far-ultraviolet (FUV) spectrum of \re
showed absorption features at the locations of the strongest \ion{Ba}{vi} and \ion{Ba}{vii} lines
as given by NIST\footnote{ National Institute of Standards and Technology,
\url{http://www.nist.gov/pml/data/asd.cfm}}. Therefore, we calculated
\ion{Ba}{v-vii} transition probabilities (Sect.\,\ref{sect:batrans}) and employed our
NLTE\footnote{non-local thermodynamic equilibrium} model-atmosphere code (Sect.\,\ref{sect:models})
to perform test calculations
(Sect.\,\ref{sect:prelim}) to find the strongest Ba lines in the model. Then, we used these
strategic lines to determine the Ba abundances of \gb and \re and searched for other, weak
Ba lines (Sect.\,\ref{sect:abund}).
We summarize our results and conclude in Sect.\,\ref{sect:results}.

\section{Transition probabilities in \ion{Ba}{v}, \ion{Ba}{vi}, and \ion{Ba}{vii}}
\label{sect:batrans}

Very few studies have been focused on the determination of electric dipole transition rates in 
\ion{Ba}{v}, \ion{Ba}{vi}, and \ion{Ba}{vii} so far. To our knowledge, the only available data were recently published 
by \citet{sharmaetal2014} and \citet{sharmaetal2013} for \ion{Ba}{v} and \ion{Ba}{vii}, respectively. More precisely, 
these authors reported oscillator strengths and transition probabilities computed using rather limited 
configuration interaction models based on the Hartree-Fock approach due to \citet{cowan1981} combined 
with a semi-empirical least-squares fit of radial energy parameters. In order to get a uniform 
set of oscillator strengths for all the transitions of Ba ions observed in the present work, we 
decided to perform new calculations including a larger amount of electron correlation and hence 
improving the previous investigations of \citet{sharmaetal2013,sharmaetal2014}. The method adopted here
was the relativistic Hartree-Fock (HFR) approach with core-polarization corrections 
\citep[see e.g.,][]{quinetetal1999,quinetetal2002}.

For \ion{Ba}{v}, configuration interaction was considered among the configurations 
5s$^2$5p$^4$, 
5s$^2$5p$^3$4f, 
5s$^2$5p$^3$5f, 
5s$^2$5p$^3$6f, 
5s$^2$5p$^3$7f, 
5s$^2$5p$^3$6p, 
5s$^2$5p$^3$7p, 
5s$^2$5p$^2$4f$^2$,
5s$^2$5p$^2$5d$^2$, 
5s$^2$5p$^2$6s$^2$, 
5s$^2$5p$^2$6p$^2$, 
5s$^2$5p$^2$5d6s, 
5s$^2$5p$^2$4f6p, 
5s5p$^4$5d, 
5s5p$^4$6d, 
5s5p$^4$6s, and
5p$^6$ for the even parity, and 
5s5p$^5$,  
5s$^2$5p$^3$5d, 
5s$^2$5p$^3$6d, 
5s$^2$5p$^3$7d, 
5s$^2$5p$^3$6s, 
5s$^2$5p$^3$7s, 
5s$^2$5p$^2$4f5d, 
5s$^2$5p$^2$4f6d, 
5s$^2$5p$^2$4f6s, 
5s$^2$5p$^2$5d6p, 
5s$^2$5p$^2$6s6p, 
5s5p$^4$4f, 
5s5p$^4$5f, 
5s5p$^4$6f, and
5s5p$^4$6p for the odd parity. 
The core-polarization parameters were the dipole polarizability of a \ion{Ba}{ix} ionic core as reported by 
\citet{fragaetal1976}, i.e., $\alpha_\mathrm{d} = 0.54$\,a.u., and the cut-off radius corresponding to the HFR 
mean value $\left<r\right>$ 
of the outermost core orbital (4d), i.e., $r_\mathrm{c} = 0.79$\,a.u. Using experimental energy levels reported by 
\citet{sharmaetal2014}, the radial integrals (average energy, Slater, spin-orbit and effective interaction 
parameters) of 
5s$^2$5p$^4$, 
5s$^2$5p$^3$6p,
5s5p$^5,$ 
5s$^2$5p$^3$5d, and 
5s$^2$5p$^3$6s configurations were optimized by a 
well-established least-squares fitting procedure in which the mean deviations with experimental data 
were found to be equal to 185\,cm$^{-1}$ for the even parity and 217\,cm$^{-1}$ for the odd parity.

For \ion{Ba}{vi}, the configurations included in the HFR model were 
5s$^2$5p$^3$, 
5s$^2$5p$^2$4f, 
5s$^2$5p$^2$5f, 
5s$^2$5p$^2$6f, 
5s$^2$5p$^2$7f, 
5s$^2$5p$^2$6p, 
5s$^2$5p$^2$7p, 
5s$^2$5p4f$^2$, 
5s$^2$5p5d$^2$, 
5s$^2$5p6s$^2$, 
5s$^2$5p6p$^2$, 
5s$^2$5p5d6s, 
5s$^2$5p4f6p, 
5s5p$^3$5d, 
5s5p$^3$6d, 
5s5p$^3$6s, and
5p$^5$ for the odd parity, and 
5s5p$^4$, 
5s$^2$5p$^2$5d, 
5s$^2$5p$^2$6d, 
5s$^2$5p$^2$7d, 
5s$^2$5p$^2$6s, 
5s$^2$5p$^2$7s, 
5s$^2$5p4f5d, 
5s$^2$5p4f6d, 
5s$^2$5p4f6s, 
5s$^2$5p5d6p, 
5s$^2$5p6s6p, 
5s5p$^3$4f, 
5s5p$^3$5f, 
5s5p$^3$6f, and 
5s5p$^3$6p for the even parity. 
In this ion, the semi-empirical process was performed to optimize the radial integrals corresponding to 
5s$^2$5p$^3$, 
5s5p$^4$, 
5s$^2$5p$^2$5d,
and 
5s$^2$5p$^2$6s configurations using the experimental levels reported by 
\citet{tauheedjoshi1994}. The mean deviations between calculated and experimental energy levels were 5\,cm$^{-1}$ 
and 128\,cm$^{-1}$ for odd and even parities, respectively. Core-polarization effects were estimated using the 
same $\alpha_\mathrm{d}$ and $r_\mathrm{c}$ values as those considered in \ion{Ba}{v}.

Finally, a similar model was used in the case of \ion{Ba}{vii} for which the 
5s$^2$5p$^2$, 
5s$^2$5p4f, 
5s$^2$5p5f, 
5s$^2$5p6f, 
5s$^2$5p7f, 
5s$^2$5p6p, 
5s$^2$5p7p, 
5s$^2$4f$^2$, 
5s$^2$5d$^2$, 
5s$^2$6s$^2$, 
5s$^2$6p$^2$, 
5s$^2$5d6s, 
5s$^2$4f6p, 
5s5p$^2$5d, 
5s5p$^2$6d, 
5s5p$^2$6s, and
5p$^4$ even-parity configurations and the 
5s5p$^3$, 
5s$^2$5p5d, 
5s$^2$5p6d, 
5s$^2$5p7d, 
5s$^2$5p6s, 
5s$^2$5p7s, 
5s$^2$4f5d, 
5s$^2$4f6d, 
5s$^2$4f6s, 
5s$^2$5d6p, 
5s$^2$6s6p, 
5s5p$^2$4f, 
5s5p$^2$5f, 
5s5p$^2$6f, and 
5s5p$^2$6p odd-parity configurations were explicitly included in the HFR model. 
Here also, we used the same core-polarization parameters as those considered for \ion{Ba}{v}. The 
semi-empirical optimization process was carried out to adjust the radial parameters in 
5s$^2$5p$^2$, 
5p$^4$, 
5s$^2$5p4f, 
5s$^2$5p6p, 
5s5p$^2$5d, 
5s5p$^3$, 
5s$^2$5p$^2$5d, and 
5s$^2$5p$^2$6s with the experimental energy levels 
classified by \citet{sharmaetal2013} below 350\,000\,cm$^{-1}$ and 320\,000\,cm$^{-1}$ for even and odd parities, respectively. 
In fact, it was found that many levels above those limits overlap unknown levels and are strongly mixed 
with states belonging to higher configurations such as 5s$^2$4f$^2$ (even parity), 5s5p$^2$4f, 5s$^2$4f5d, and 5s$^2$5p6d 
(odd parity). It was then extremely difficult to establish an unambiguous correspondence between the 
calculated values and the experimentally determined level energies. The use of radial parameters 
published by \citet{sharmaetal2013} was unfortunately not of great help for making the identifications more 
reliable, the set of interacting configurations being not the same as the one considered in our work. 
For the levels considered in our fitting process, the mean deviations between calculated and 
experimental values were found to be equal to 140\,cm$^{-1}$ (even parity) and 196\,cm$^{-1}$ (odd parity). 

\begin{figure*}
   \sidecaption
   \includegraphics[width=12cm]{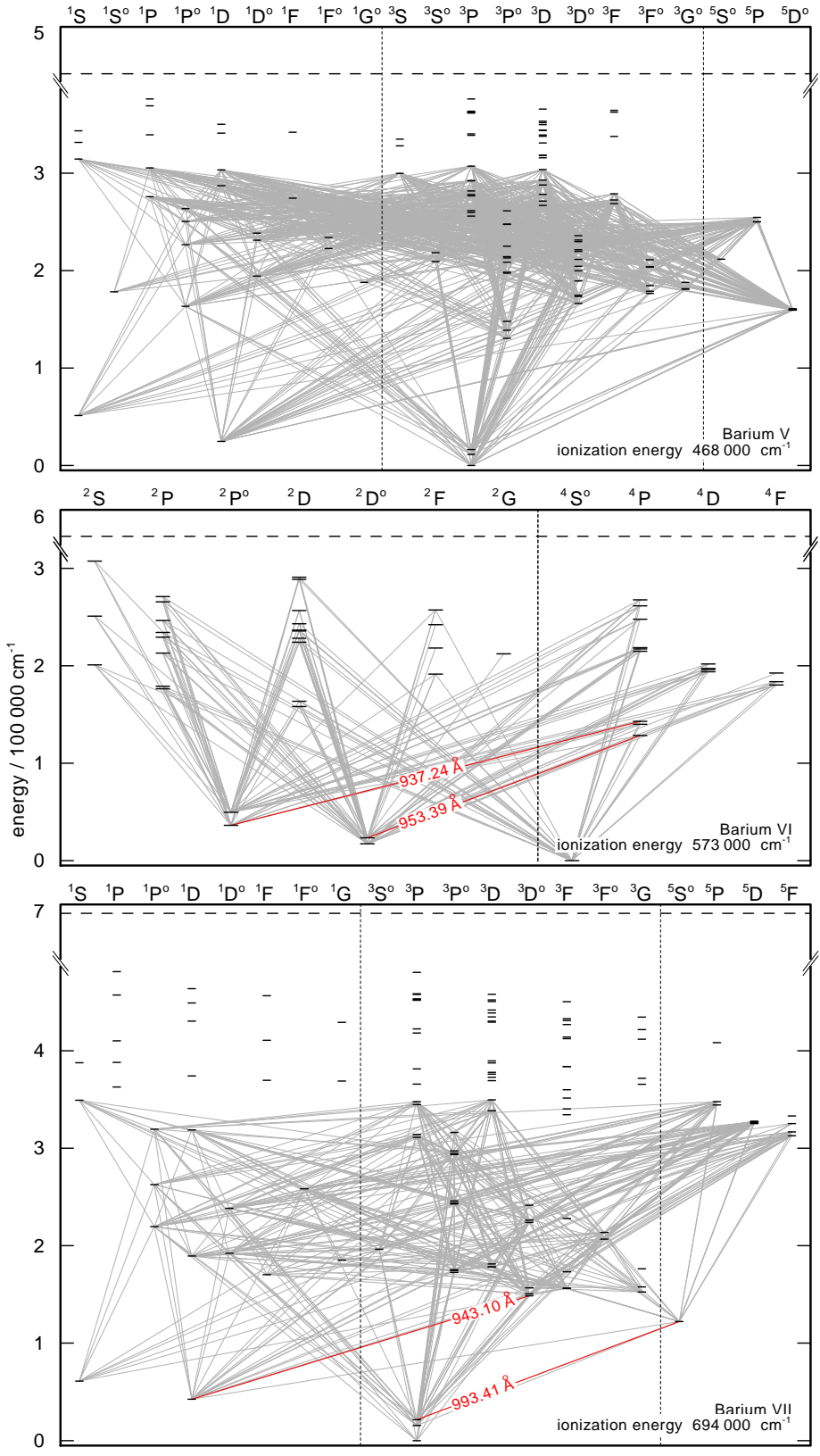}
      \caption{Grotrian diagrams of our 
               \ion{Ba}{v} (top),
               \ion{Ba}{vi} (middle), and
               \ion{Ba}{vii} (bottom) model ions.
               Horizontal bars indicate levels,
               gray lines represent radiative
               transitions with known $f$ values
               (Tables \ref{tab:bav:loggf}, \ref{tab:bavi:loggf}, and \ref{tab:bavii:loggf}).
               Our strategic \ion{Ba}{vi} and \ion{Ba}{vii} lines 
               in the FUSE wavelength range (Fig.\,\ref{fig:abund}) are
               labeled with their wavelengths.
               The long-dashed, horizontal lines in each panel show the ionization energies
               of 468\,000\,cm$^{-1}$, 573\,000\,cm$^{-1}$, and 694\,000\,cm$^{-1}$ of
               \ion{Ba}{v}, \ion{Ba}{vi}, and \ion{Ba}{vii}, respectively.
              }
   \label{fig:grotrian}
\end{figure*}

Figure\,\ref{fig:grotrian} shows Grotrian diagrams of \ion{Ba}{v-vii} including
all levels and transitions from Tables \ref{tab:bav:loggf}, \ref{tab:bavi:loggf}, and \ref{tab:bavii:loggf}.

\onltab{
\onecolumn
\begin{longtable}{rrccrrccrrr}
\caption{\label{tab:bav:loggf}Calculated HFR oscillator strengths ($\log gf$) and transition probabilities (gA, in sec$^{-1}$) 
                              in \ion{Ba}{v}. CF is the cancellation factor as defined by \citet{cowan1981}. 
                              In columns 3 and 6, e is written for even and o for odd.} \\
\hline\hline
\noalign{\smallskip}
                 & \multicolumn{3}{c}{Lower level} && \multicolumn{3}{c}{Upper level} &           &      &    \\
\cline{2-4}\cline{6-8}
\noalign{\smallskip}
Wavelength / \AA & Energy / cm$^{-1}$ & Parity & $j$  && Energy / cm$^{-1}$ & Parity & $j$  & $\log gf$ & $gA$ & CF \\
\noalign{\smallskip}
\hline
\endfirsthead
\caption{continued.}\\
\hline\hline
\noalign{\smallskip}
                 & \multicolumn{3}{c}{Lower level} && \multicolumn{3}{c}{Upper level} &           &      &    \\
\cline{2-4}\cline{6-8}
\noalign{\smallskip}
Wavelength / \AA & Energy / cm$^{-1}$ & Parity & $j$  && Energy / cm$^{-1}$ & Parity & $j$  & $\log gf$ & $gA$ & CF \\
\noalign{\smallskip}
\hline
\noalign{\smallskip}
\endhead
\hline
\noalign{\smallskip}
\endfoot
\noalign{\smallskip}
     382.885 &         0 & e & 2.0 & &    261175 & o &  2.0 &$-$2.12 &  $3.50 \times 10^{08} $ & $-$0.004 \\
     396.365 &     11302 & e & 0.0 & &    263594 & o &  1.0 &$-$2.78 &  $6.99 \times 10^{07} $ &    0.001 \\
     399.297 &         0 & e & 2.0 & &    250440 & o &  1.0 &$-$2.81 &  $6.49 \times 10^{07} $ &    0.001 \\
     404.183 &         0 & e & 2.0 & &    247413 & o &  1.0 &$-$2.81 &  $6.26 \times 10^{07} $ & $-$0.001 \\
     404.588 &     16429 & e & 1.0 & &    263594 & o &  1.0 &$-$1.51 &  $1.27 \times 10^{09} $ & $-$0.045 \\
     408.587 &     16429 & e & 1.0 & &    261175 & o &  2.0 &$-$0.44 &  $1.47 \times 10^{10} $ &    0.395 \\
     418.168 &     11302 & e & 0.0 & &    250440 & o &  1.0 &$-$0.65 &  $8.48 \times 10^{09} $ & $-$0.039 \\
     418.631 &     24721 & e & 2.0 & &    263594 & o &  1.0 &$-$0.39 &  $1.55 \times 10^{10} $ & $-$0.263 \\
\hline
\end{longtable}
\twocolumn
}

\onltab{
\onecolumn
\begin{longtable}{rrccrrccrrr}
\caption{\label{tab:bavi:loggf}Like Table\,\ref{tab:bav:loggf}, for \ion{Ba}{vi}.} \\
\hline\hline
\noalign{\smallskip}
                 & \multicolumn{3}{c}{Lower level} && \multicolumn{3}{c}{Upper level} &           &      &    \\
\cline{2-4}\cline{6-8}
\noalign{\smallskip}
Wavelength / \AA & Energy / cm$^{-1}$ & Parity & $j$  && Energy / cm$^{-1}$ & Parity & $j$  & $\log gf$ & $gA$ & CF \\
\noalign{\smallskip}
\hline
\endfirsthead
\caption{continued.}\\
\hline\hline
\noalign{\smallskip}
                 & \multicolumn{3}{c}{Lower level} && \multicolumn{3}{c}{Upper level} &           &      &    \\
\cline{2-4}\cline{6-8}
\noalign{\smallskip}
Wavelength / \AA & Energy / cm$^{-1}$ & Parity & $j$  && Energy / cm$^{-1}$ & Parity & $j$  & $\log gf$ & $gA$ & CF \\
\noalign{\smallskip}
\hline
\noalign{\smallskip}
\endhead
\hline
\noalign{\smallskip}
\endfoot
\noalign{\smallskip}
     343.756 &         0 & o & 1.5 & &    290904 & e &  1.5 &$-$3.19 &  $3.61 \times 10^{07} $ &    0.001 \\
     344.677 &     17261 & o & 1.5 & &    307387 & e &  0.5 &$-$2.88 &  $7.42 \times 10^{07} $ & $-$0.003 \\
     346.542 &         0 & o & 1.5 & &    288565 & e &  2.5 &$-$1.75 &  $9.91 \times 10^{08} $ &    0.014 \\
     365.440 &     17261 & o & 1.5 & &    290904 & e &  1.5 &$-$0.81 &  $7.66 \times 10^{09} $ & $-$0.069 \\
     368.590 &     17261 & o & 1.5 & &    288565 & e &  2.5 &$-$0.29 &  $2.50 \times 10^{10} $ &    0.608 \\
     368.689 &     36156 & o & 0.5 & &    307387 & e &  0.5 &$-$0.48 &  $1.64 \times 10^{10} $ & $-$0.538 \\
     368.890 &         0 & o & 1.5 & &    271083 & e &  1.5 &$-$1.02 &  $4.73 \times 10^{09} $ & $-$0.082 \\
     373.623 &         0 & o & 1.5 & &    267650 & e &  2.5 &$-$0.04 &  $4.37 \times 10^{10} $ &    0.510 \\
\hline
\end{longtable}
\twocolumn
}

\onltab{
\onecolumn
\begin{longtable}{rrccrrccrrr}
\caption{\label{tab:bavii:loggf}Like Table\,\ref{tab:bav:loggf}, for \ion{Ba}{vii}.} \\
\hline\hline
\noalign{\smallskip}
                 & \multicolumn{3}{c}{Lower level} && \multicolumn{3}{c}{Upper level} &           &      &    \\
\cline{2-4}\cline{6-8}
\noalign{\smallskip}
Wavelength / \AA & Energy / cm$^{-1}$ & Parity & $j$  && Energy / cm$^{-1}$ & Parity & $j$  & $\log gf$ & $gA$ & CF \\
\noalign{\smallskip}
\hline
\endfirsthead
\caption{continued.}\\
\hline\hline
\noalign{\smallskip}
                 & \multicolumn{3}{c}{Lower level} && \multicolumn{3}{c}{Upper level} &           &      &    \\
\cline{2-4}\cline{6-8}
\noalign{\smallskip}
Wavelength / \AA & Energy / cm$^{-1}$ & Parity & $j$  && Energy / cm$^{-1}$ & Parity & $j$  & $\log gf$ & $gA$ & CF \\
\noalign{\smallskip}
\hline
\noalign{\smallskip}
\endhead
\hline
\noalign{\smallskip}
\endfoot
\noalign{\smallskip}
     312.932 &         0 & e & 0.0 & &    319558 & o &  1.0 &$-$2.48 &  $2.27 \times 10^{08} $ &$-$0.009 \\
     328.892 &     15507 & e & 1.0 & &    319558 & o &  1.0 &$-$1.15 &  $4.40 \times 10^{09} $ &   0.747 \\
     332.572 &     15507 & e & 1.0 & &    316193 & o &  2.0 &$-$0.39 &  $2.46 \times 10^{10} $ &$-$0.635 \\
     335.504 &     21499 & e & 2.0 & &    319558 & o &  1.0 &$-$0.59 &  $1.54 \times 10^{10} $ &$-$0.189 \\
     339.116 &         0 & e & 0.0 & &    294885 & o &  1.0 &$-$0.41 &  $2.27 \times 10^{10} $ &$-$0.700 \\
     339.335 &     21499 & e & 2.0 & &    316193 & o &  2.0 &$-$0.09 &  $4.68 \times 10^{10} $ &   0.762 \\
     357.938 &     15507 & e & 1.0 & &    294885 & o &  1.0 &$-$0.72 &  $9.99 \times 10^{09} $ &   0.725 \\
     359.650 &     15507 & e & 1.0 & &    293555 & o &  0.0 &$-$0.48 &  $1.71 \times 10^{10} $ &   0.774 \\
\hline
\end{longtable}
\twocolumn
}

\section{Observations}
\label{sect:observation}

In this analysis, we use the 
FUSE\footnote{Far Ultraviolet Spectroscopic Explorer} spectrum of \re and 
the FUSE and HST/STIS\footnote{Hubble Space Telescope / Space Telescope Imaging Spectrograph,
                  for our high-resolution spectrum of \gb, see \url{http://www.stsci.edu/hst/observatory/crds/calspec.html}} 
spectra of \gb that are described in detail by \citet{werneretal2012} and \citet{rauchetal2013,rauchetal2014}, respectively.

The FUSE spectrum covers the wavelength range $910\,\mathrm{\AA} < \lambda <  1188\,\mathrm{\AA}$. 
Its resolving power is $R = \lambda/\Delta\lambda \approx 20\,000$.
The HST/STIS spectrum of \gb is co-added from 105 observations with the highest resolution (grating E140H,
$R \approx 118\,000$, $1145\,\mathrm{\AA} < \lambda < 1750\,\mathrm{\AA}$) available via MAST.

\section{Model atmospheres and atomic data}
\label{sect:models}

Our model atmospheres (plane-parallel, chemically homogeneous, in radiative and hydrostatic equilibrium)
were calculated with the 
T\"ubingen NLTE model-atmosphere package 
\citep[TMAP\footnote{\url{http://astro.uni-tuebingen.de/~TMAP}},][]{werneretal2003,rauchdeetjen2003}.

The model atoms (including Ba, Table\,\ref{tab:statistics} displays its statistics) that are considered 
in our calculations are provided by the
T\"ubingen Model-Atom Database
\citep[TMAD\footnote{\url{http://astro.uni-tuebingen.de/~TMAD}},][]{rauchdeetjen2003},
that was constructed in the framework of the German Astrophysical Virtual Observatory 
(GAVO\footnote{\url{http://www.g-vo.org}}).
The SEDs that were calculated for this analysis are available via
the registered Theoretical Stellar Spectra Access 
(TheoSSA\footnote{\url{http://dc.g-vo.org/theossa}}) VO service.

\begin{table}\centering
\caption{Statistics of our Ba model atoms.}
\label{tab:statistics}
\begin{tabular}{r@{~}lrrr}
\hline\hline
\noalign{\smallskip}
\multicolumn{2}{c}{ion~~~~}  & NLTE levels &   LTE levels &  lines \\
\noalign{\smallskip}                                                        
\hline                                                                      
\noalign{\smallskip}                                                  
Ba & {\sc iv}   &      1\hspace{5mm}\hbox{} &   6\hspace{4mm}\hbox{} &    0\hspace{2mm}\hbox{} \\
   & {\sc v}    &    123\hspace{5mm}\hbox{} &   0\hspace{4mm}\hbox{} &  981\hspace{2mm}\hbox{} \\
   & {\sc vi}   &     47\hspace{5mm}\hbox{} &   0\hspace{4mm}\hbox{} &  162\hspace{2mm}\hbox{} \\
   & {\sc vii}  &    121\hspace{5mm}\hbox{} &   0\hspace{4mm}\hbox{} &  452\hspace{2mm}\hbox{} \\
   & {\sc viii} &      1\hspace{5mm}\hbox{} &   0\hspace{4mm}\hbox{} &    0\hspace{2mm}\hbox{} \\
\hline
\end{tabular}
\end{table}

The start models for our calculations are the most elaborated model atmospheres for both stars \citep{rauchetal2014}
As our main model-atmosphere program would not compile if the array sizes were increased to accommodate 
the high number of atomic levels treated in NLTE and the high number of radiative and collisional transitions,
\citep[cf\@.][]{rauchetal2014}, 
we simply reduced the number of the Zn levels treated in NLTE to one per ion to create a TMAP executable. 
Because the Zn opacities were already considered in detail in
our start models, the atmospheric structure and the background opacities are well modeled. 
To calculate the NLTE occupation numbers of Ba, we performed line-formation calculations, i.e., we kept
temperature and density structure of our model-atmospheres fixed. Since the impact of Ba
on the atmospheric structure of our H+Ba and He+Ba test models (Sect.\,\ref{sect:prelim}) was found to be marginal, this
is the best practice.  For the subsequent SED calculations, we considered the complete model ions 
and resumed the missing Zn occupation numbers from the start models.

\section{Preliminary analysis}
\label{sect:prelim}

\begin{figure}
   \resizebox{\hsize}{!}{\includegraphics{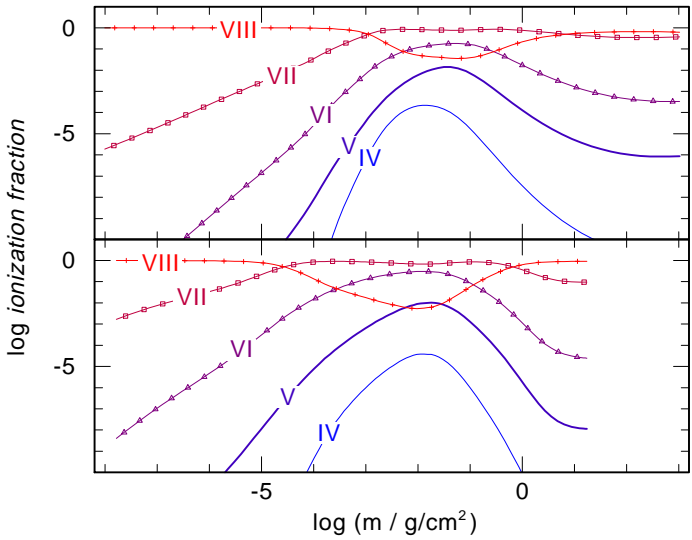}}
    \caption{Ba ionization fractions in our models for 
             \gb (top panel, H+Ba) and
             \re (bottom panel, He+Ba).
             $m$ is the column mass, measured from the outer boundary of our model atmospheres.
            }
   \label{fig:ion}
\end{figure}

\begin{figure}
   \resizebox{\hsize}{!}{\includegraphics{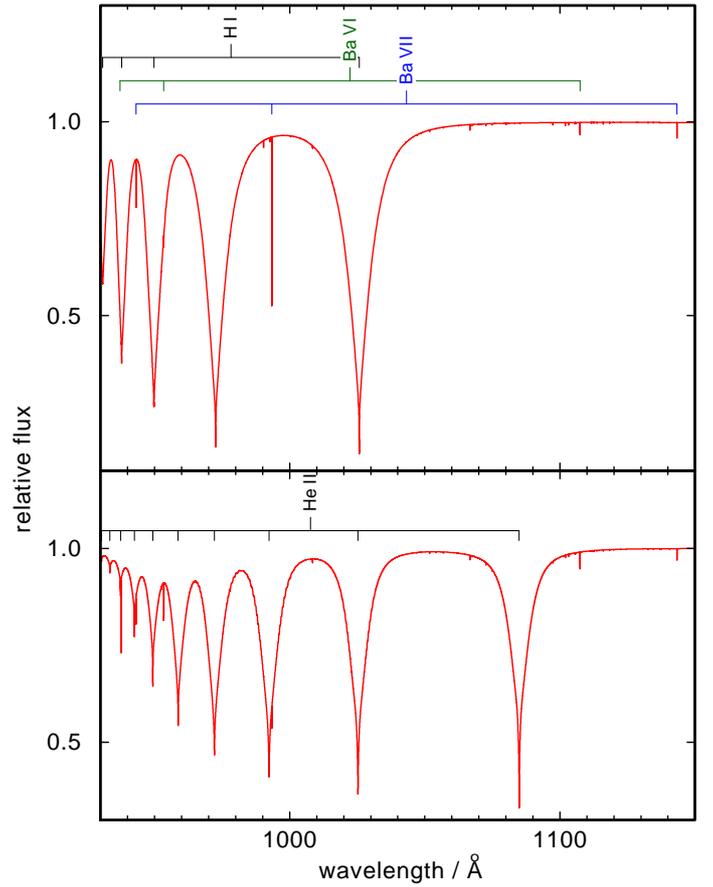}}
    \caption{Synthetic spectra (normalized to the models's continuum fluxes) of 
             our H+Ba model for \gb (top panel) and 
             our He+Ba model for \re (bottom panel).
             All synthetic spectra displayed in this paper are convolved
             with Gaussians to simulate the instruments' resolution,
             here, for FUSE, we used an FWHM (full width at half maximum) of 0.06\,\AA.
            }
   \label{fig:HBa_FUSE}
\end{figure}

For a preliminary analysis, we calculated models that consider only H+Ba and He+Ba for
\gb (\Teffw{60\,000}, \loggw{7.6}) and 
\re (\Teffw{70\,000}, \loggw{7.5}), 
respectively. The Ba abundance was $1.5 \times 10^{-5}$
\citep[mass fraction, about 1000 times solar,][]{asplundetal2009}
in both models. Figure\,\ref{fig:ion} shows
the respective Ba ionization fractions. 
\ion{Ba}{vii} is the dominating ionization stage in the line
forming region ($-2.5\,\sla\,\log m\,\sla\,0.5$). In the case of \gb, the H+Ba spectrum shows two prominent 
lines in the FUV, \Ionww{Ba}{7}{943.10, 993.41}, and several much weaker \ion{Ba}{vi} and 
\ion{Ba}{vii} lines (Fig.\,\ref{fig:HBa_FUSE}).
For \re, \Ionw{Ba}{7}{993.41} is the strongest line in the He+Ba, but some weaker \ion{Ba}{vi} and 
\ion{Ba}{vii} lines are also visible.

\begin{figure}
   \resizebox{\hsize}{!}{\includegraphics{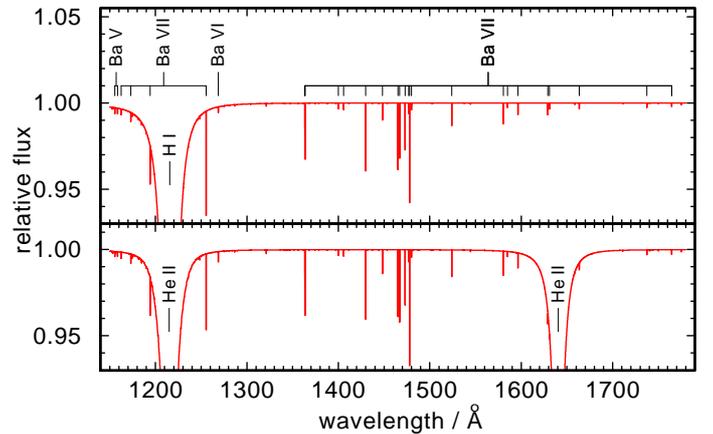}}
    \caption{Same as Fig.\,\ref{fig:HBa_FUSE} (top: \gb, bottom: \re), 
             to simulate the HST/STIS spectrum (FWHM = 0.01\,\AA).
            }
   \label{fig:HBa_STIS}
\end{figure}

For the wavelengths higher than the FUSE wavelength range, an adequate observation is only available
for \gb (Sect.\,\ref{sect:observation}). Our test model shows a rich \ion{Ba}{vii} spectrum within
$1150\,\mathrm{\AA} \le \lambda \le 1780\,\mathrm{\AA}$ for this star, and a few weak \ion{Ba}{v-vi}
lines in addition (Fig.\,\ref{fig:HBa_STIS}). The model spectrum for \re shows the same lines,
with deviations in the relative line strengths. We note that all these lines are much weaker than the four
\ion{Ba}{vi-vii} lines in the FUSE wavelength range (Fig.\,\ref{fig:HBa_FUSE}).

\begin{figure}
   \resizebox{\hsize}{!}{\includegraphics{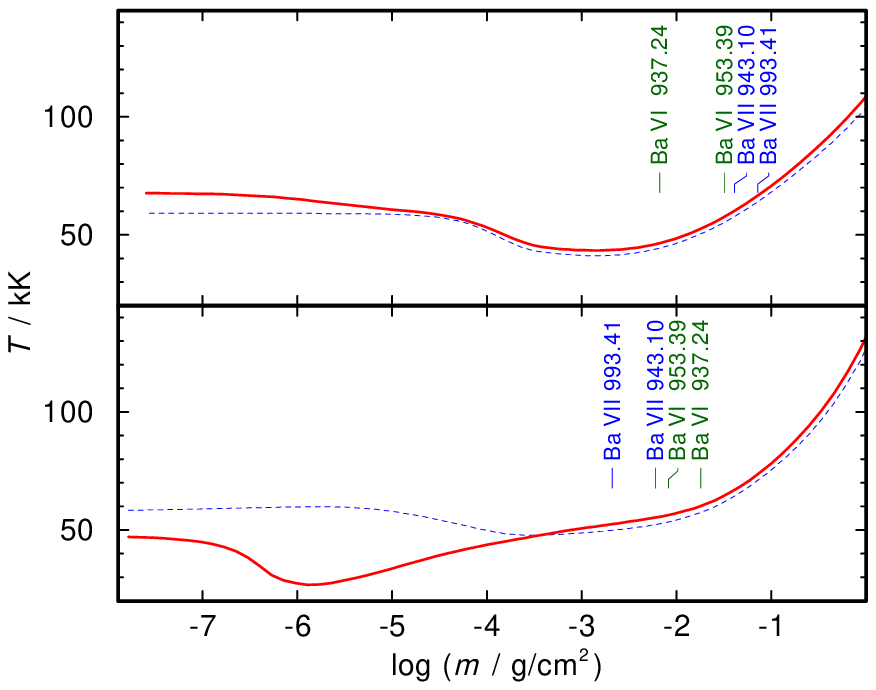}}
    \caption{Temperature structures of the final model atmospheres of 
             \gb 
             \citep[top panel, thick, red line;]
                   [H, He, C, N, O, Al, Si, P, S, Ca, Sc, Ti, V, Cr, Mn, Fe, Co, Ni, Zn, and Ge opacities considered]
                   {rauchetal2013}
             and \re 
             \citep[bottom panel, thick, red line;]
                   [He, C, N, O, Ca, Sc, Ti, V, Cr, Mn, Fe, Co, Ni, Zn, Ge, Kr, and Xe]
                   {rauchetal2014}. 
             The dashed, blue lines show the temperature structures of H+Ba (top) and He+Ba (bottom) models
             calculated with the same \Teff and \logg like the final models, respectively.
             The formation depths 
             (i.e\@., $\tau = 1$) 
             of the line cores of our strategic \ion{Ba}{vi} - {\sc vii} in FUSE's wavelength range are marked.
            }
   \label{fig:formdep}
\end{figure}

The preliminary models cannot be used for a reliable abundance determination because
the neglected metals result in an unrealistic atmospheric structure (Fig.\,\ref{fig:formdep}),
and their missing background opacities have a strong impact on the strengths of the Ba lines.
Therefore, we performed a precise determination of the Ba abundances
based on detailed atomic data and elaborated model atmospheres (Sect.\,\ref{sect:abund}).

\section{The photospheric Ba abundances in \gb and \re}
\label{sect:abund}

\begin{figure}
   \resizebox{\hsize}{!}{\includegraphics{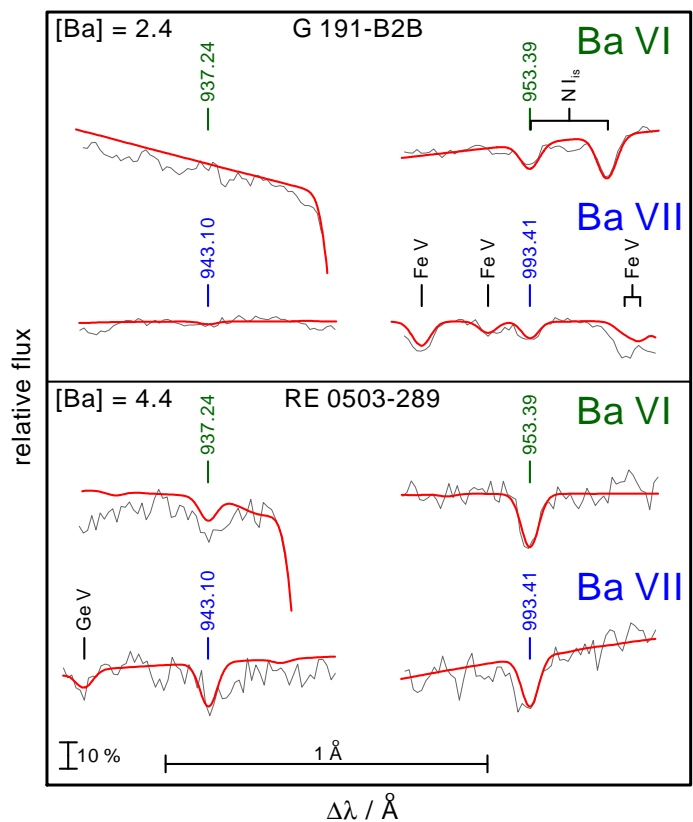}}
    \caption{Comparison of the theoretical line profiles of the strongest
             \ion{Ba}{vi} and \ion{Ba}{vii} lines with the FUSE observation.
             The synthetic spectra were normalized to the observed local continuum fluxes. 
             Top: \gb, bottom: \re.
             [X] denotes log ( mass fraction / solar mass fraction ) of species X.
            }
   \label{fig:abund}
\end{figure}

\begin{figure}
   \resizebox{\hsize}{!}{\includegraphics{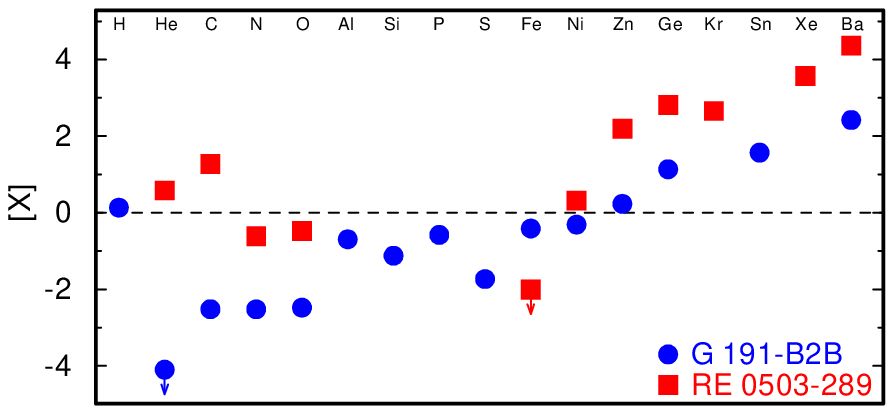}}
    \caption{Photospheric abundances of 
             \gb \citep[bullets,][]{rauchetal2013,rauchetal2014} and 
             \re \citep[squares,][]{werneretal2012,rauchetal2012,rauchetal2014} compared with solar values
             \citep{asplundetal2009}. The abundance uncertainties are about 0.2\,dex in general. The
             arrows denote upper limits.
             The dashed line indicates the solar values.
            }
   \label{fig:abundpattern}
\end{figure}

We determine the Ba abundances of \gb and \re from their FUSE
observations. Figure\,\ref{fig:abund} shows that we can well reproduce the Ba lines 
with Ba mass fractions of $4.0 \pm 0.5 \times 10^{-6}$ and $3.5 \pm 0.5 \times 10^{-4}$,
respectively. These values are strongly over-solar \citep[23\,000 times and 265 times,
respectively,][]{asplundetal2009} but in line with the determined abundances of other trans-iron
elements \citep[e.g\@.,][Fig.\,\ref{fig:abundpattern}]{werneretal2012,rauchetal2013}. It is worthwhile to note,
that the \ion{Ba}{vi} / \ion{Ba}{vii} ionization equilibrium is very well reproduced for \re.

\begin{figure}
   \resizebox{\hsize}{!}{\includegraphics{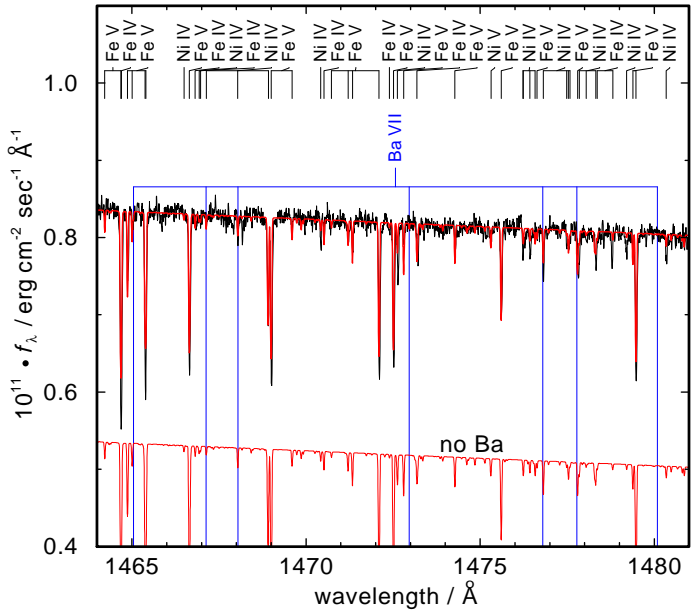}}
    \caption{Section of the HST/STIS spectrum of \gb compared with our synthetic
             spectrum. The thin (blue) model at the bottom is calculated without Ba.
             The identified Fe and Ni lines and the \ion{Ba}{vii} lines are marked at top.}
   \label{fig:bavii_STIS}
\end{figure}

In addition to the metals' background opacity in the final model mentioned above,
the determined Ba abundance of \gb is only a fourth of that used in our test model (Fig.\,\ref{fig:HBa_STIS})
and, thus, the lines in the wavelength interval $1145\,\mathrm{\AA} < \lambda < 1750\,\mathrm{\AA}$ 
are weaker in the models, respectively. Figure\,\ref{fig:bavii_STIS}
shows a section of the spectrum where our test models predicted the strongest lines (Fig.\,\ref{fig:HBa_STIS}).
A comparison to a model, that was calculated without Ba, shows that in this section only 
\Ionw{Ba}{7}{1472.96} is a weak, isolated line that is visible in our model. Its strength is comparable to the
low noise of the observed spectrum and, thus, we can determine an upper limit for the Ba abundance 
which is the same upper value determined from the FUSE spectrum (see above). 
For \re, the determined Ba abundance is more than 20 times higher than in our test model (Fig.\,\ref{fig:HBa_STIS})
and, thus, the potential identification of Ba lines in this wavelength range is most likely. A high-resolution,
high-signal-to-noise HST/STIS UV spectrum of \re is highly desirable.

\section{Results and conclusions}
\label{sect:results}

We determined precisely the photospheric Ba abundances 
in the DA-type white dwarf \gb ($4.0 \pm 0.5 \times 10^{-6}$) and 
in the DO-type white dwarf \re ($3.5 \pm 0.5 \times 10^{-4}$).
These strongly supersolar abundance values are in line with those
of other trans-iron elements in both stars (Fig.\,\ref{fig:abundpattern}).

Ba is one of the most massive s-process (slow neutron capture) elements synthesized by 
low-mass ($\approx 1 - 3$\,\Msol) AGB stars \citep[e.g.,][]{lattanziolugaro2005}. 
The s-process leads to abundance peaks at approximate mass numbers of 88, 138, and 208 
\citep{westheger2013,karakasetal2014} due to closed neutron shells. Sr, Ba (isotopes from $^{128}$Ba to 
$^{140}$Ba), and Pb are located at these peaks and are used to represent the scaling of the
s-process elements.
Therefore, the Ba abundance determined in the hot WDs \gb and \re 
establishes a new constraint for AGB and post-AGB stellar evolution
and will help to understand the extremely strong metal enrichment.

The identification of Ba lines in this paper was only possible because reliable 
transition probabilities for \ion{Ba}{v}, \ion{Ba}{vi}, and \ion{Ba}{vii}
were calculated.
Analogous calculations for other highly ionized trans-iron elements is a
pre-requisite for further identifications an abundance analyses. 
The precise measurement of their spectra, i.e., their lines' wavelengths and relative strengths,
as well as the determination of level energies and the calculation of transition probabilities 
remains a challenge for atomic and theoretical physicists.

\begin{acknowledgements}
TR is supported by the German Aerospace Center (DLR, grant 05\,OR\,1301).
Financial support from the Belgian FRS-FNRS is also acknowledged. 
PQ is research director of this organization.
This research has made use of the SIMBAD database, operated at CDS, Strasbourg, France.
Some of the data presented in this paper were obtained from the
Mikulski Archive for Space Telescopes (MAST). STScI is operated by the
Association of Universities for Research in Astronomy, Inc., under NASA
contract NAS5-26555. Support for MAST for non-HST data is provided by
the NASA Office of Space Science via grant NNX09AF08G and by other
grants and contracts. 

\end{acknowledgements}

\bibliographystyle{aa}
\bibliography{23878}

\begin{thebibliography}{17}
\expandafter\ifx\csname natexlab\endcsname\relax\def\natexlab#1{#1}\fi

\bibitem[{{Asplund} {et~al.}(2009){Asplund}, {Grevesse}, {Sauval}, \&
  {Scott}}]{asplundetal2009}
{Asplund}, M., {Grevesse}, N., {Sauval}, A.~J., \& {Scott}, P. 2009, \araa, 47,
  481

\bibitem[{{Cowan}(1981)}]{cowan1981}
{Cowan}, R.~D. 1981, {The theory of atomic structure and spectra} (Berkeley,
  CA, University of California Press)

\bibitem[{{Fraga} {et~al.}(1976){Fraga}, {Karwowski}, \&
  {Saxena}}]{fragaetal1976}
{Fraga}, S., {Karwowski}, J., \& {Saxena}, K.~M.~S. 1976, {Handbook of Atomic
  Data} (Elsevier, Amsterdam)

\bibitem[{{Karakas} {et~al.}(2014){Karakas}, {Marino}, \&
  {Nataf}}]{karakasetal2014}
{Karakas}, A.~I., {Marino}, A.~F., \& {Nataf}, D.~M. 2014, \apj, 784, 32

\bibitem[{{Lattanzio} \& {Lugaro}(2005)}]{lattanziolugaro2005}
{Lattanzio}, J.~C. \& {Lugaro}, M.~A. 2005, Nuclear Physics A, 758, 477

\bibitem[{{Quinet} {et~al.}(2002){Quinet}, {Palmeri}, {Bi{\'e}mont}, {Li},
  {Zhang}, \& Svanberg{}}]{quinetetal2002}
{Quinet}, P., {Palmeri}, P., {Bi{\'e}mont}, E., {et~al.} 2002, in J. Alloys
  Comp., Vol. 344, Proceedings of the Rare Earths` 2001 Conference, ed. M.~A.
  {Leskela}, M.~F. {Reid}, H.~B. {Silber}, \& L.~B. {Zinner}, 255

\bibitem[{{Quinet} {et~al.}(1999){Quinet}, {Palmeri}, {Bi{\'e}mont}, {McCurdy},
  {Rieger}, {Pinnington}, {Wickliffe}, \& {Lawler}}]{quinetetal1999}
{Quinet}, P., {Palmeri}, P., {Bi{\'e}mont}, E., {et~al.} 1999, \mnras, 307, 934

\bibitem[{{Rauch} \& {Deetjen}(2003)}]{rauchdeetjen2003}
{Rauch}, T. \& {Deetjen}, J.~L. 2003, in Astronomical Society of the Pacific
  Conference Series, Vol. 288, Stellar Atmosphere Modeling, ed. I.~{Hubeny},
  D.~{Mihalas}, \& K.~{Werner}, 103

\bibitem[{{Rauch} {et~al.}(2012){Rauch}, {Werner}, {Bi{\'e}mont}, {Quinet}, \&
  {Kruk}}]{rauchetal2012}
{Rauch}, T., {Werner}, K., {Bi{\'e}mont}, {\'E}., {Quinet}, P., \& {Kruk},
  J.~W. 2012, \aap, 546, A55

\bibitem[{{Rauch} {et~al.}(2013){Rauch}, {Werner}, {Bohlin}, \&
  {Kruk}}]{rauchetal2013}
{Rauch}, T., {Werner}, K., {Bohlin}, R., \& {Kruk}, J.~W. 2013, \aap, 560, A106

\bibitem[{{Rauch} {et~al.}(2014){Rauch}, {Werner}, {Quinet}, \&
  {Kruk}}]{rauchetal2014}
{Rauch}, T., {Werner}, K., {Quinet}, P., \& {Kruk}, J.~W. 2014, \aap, 564, A41

\bibitem[{{Sharma} {et~al.}(2014){Sharma}, {Rahimullah}, \&
  {Tauheed}}]{sharmaetal2014}
{Sharma}, M.~K., {Rahimullah}, K., \& {Tauheed}, A. 2014, \jqsrt, 133, 281

\bibitem[{{Sharma} {et~al.}(2013){Sharma}, {Tauheed}, \&
  {Rahimullah}}]{sharmaetal2013}
{Sharma}, M.~K., {Tauheed}, A., \& {Rahimullah}, K. 2013, \jqsrt, 119, 32

\bibitem[{{Tauheed} \& {Joshi}(1994)}]{tauheedjoshi1994}
{Tauheed}, A. \& {Joshi}, Y.~N. 1994, \physscr, 49, 335

\bibitem[{{Werner} {et~al.}(2003){Werner}, {Deetjen}, {Dreizler}, {Nagel},
  {Rauch}, \& {Schuh}}]{werneretal2003}
{Werner}, K., {Deetjen}, J.~L., {Dreizler}, S., {et~al.} 2003, in Astronomical
  Society of the Pacific Conference Series, Vol. 288, Stellar Atmosphere
  Modeling, ed. I.~{Hubeny}, D.~{Mihalas}, \& K.~{Werner}, 31

\bibitem[{{Werner} {et~al.}(2012){Werner}, {Rauch}, {Ringat}, \&
  {Kruk}}]{werneretal2012}
{Werner}, K., {Rauch}, T., {Ringat}, E., \& {Kruk}, J.~W. 2012, \apjl, 753, L7

\bibitem[{{West} \& {Heger}(2013)}]{westheger2013}
{West}, C. \& {Heger}, A. 2013, \apj, 774, 75

\end{thebibliography}

\end{document}